\documentclass[pdflatex]{article}
\usepackage{dcase2019,amsmath,times}
\usepackage{graphicx}
\usepackage{bm}
\usepackage{url}
\usepackage{amssymb}
\usepackage{cite}

\usepackage{multirow}
\usepackage{makecell}
\usepackage{array}
\usepackage{algorithmicx}

\makeatletter
\def\Hline{
  \noalign{\ifnum0=`}\fi\hrule \@height 3.\arrayrulewidth \futurelet
  \reserved@a\@xhline}
\makeatother

\title{FIRST ORDER AMBISONICS DOMAIN SPATIAL AUGMENTATION FOR DNN-BASED DIRECTION OF ARRIVAL ESTIMATION}

 \name{Luca Mazzon${}^{1,2}$, Yuma Koizumi${}^{1}$, Masahiro Yasuda${}^{1}$, and Noboru Harada${}^{1}$}
\address{
${}^{1}$NTT Media Intelligence Laboratories, Tokyo, Japan\\
${}^{2}$University of Padova, Padua, Italy
}

\begin{document}

\ninept
\maketitle

\begin{sloppy}

\begin{abstract}
In this paper, we propose a novel data augmentation method for training neural networks for Direction of Arrival (DOA) estimation. This method focuses on expanding the representation of the DOA subspace of a dataset. Given some input data, it applies a transformation to it in order to change its DOA information and simulate new potentially unseen one. Such transformation, in general, is a combination of a rotation and a reflection. It is possible to apply such transformation due to a well-known property of First Order Ambisonics (FOA). The same transformation is applied also to the labels, in order to maintain consistency between input data and target labels. Three methods with different level of generality are proposed for applying this augmentation principle. Experiments are conducted on two different DOA networks. Results of both experiments demonstrate the effectiveness of the novel augmentation strategy by improving the DOA error by around 40\%.
\end{abstract}

\begin{keywords}
First Order Ambisonics, direction of arrival, deep learning, data augmentation
\end{keywords}

\section{Introduction}
\label{sec:intro}

Direction of arrival (DOA) estimation is the task of detecting the spatial position of a sound source with respect to a listener. The approaches that has been adopted to solve this problem can be classified in two main categories: parametric-based methods, like multiple signal classification (MUSIC) \cite{MUSIC} and others \cite{Huang, Brandstein, Roy}, and deep neural network (DNN)-based methods \cite{DOAnet, SELDnet, CaoIqbal, Hirvonen, Yiwere, Ferguson, Vesperini, Sun, Roden,
DiBiase, Takeda, Yalta, He
}. 
DNN-based models often combine DOA estimation with other tasks such as sound activity detection (SAD), estimation of number of active sources and sound event detection (SED) \cite{Hirvonen, SELDnet, CaoIqbal}. In particular, Sound Event Localization and Detection was the task 3 of Detection and Classification of Acoustic Scenes and Events 2019 Challenge (DCASE2019 Challenge) \cite{DCASE2019Task3}.

In machine learning, \textit{data augmentation} is an effective strategy to overcome the lack of data in the training set and prevent overfitting. For example SpecAugment \cite{SpeckAugment}, a recently published augmentation method based on time warping and time and frequency block masking of the spectrogram, achieved state of the art performance on the Speech recognition task. DCASE2018 Task2 Challenge (about audio tagging) winner \cite{Jeong2018} used mixup augmentation \cite{mixup}.

While data augmentation is effective for sound event detection and similar tasks, none of the documented strategies is capable of effectively increasing the spatial representativeness of a dataset, i.e. increasing the number of DOAs represented in the dataset.
The critical point of the problem is that when the observed signals are modified by a data augmentation method, it must be guaranteed that the relationship between the DOA information carried by the signal and the corresponding labels is maintained. For example, augmentation techniques such as SpecAugment, phase-shifting and mixup can indeed influence DOA, although it's hard to analytically compute the new true DOA labels.
In fact, according to the technical reports of DCASE 2019 task3, SpecAugment has affected adversely for DOA estimation even though it is effective for SED \cite{Kapka,task3_he}.

In this paper, we propose \textit{FOA Domain Spatial Augmentation}, a novel augmentation method based on the well-known rotational property of First Order Ambisonics (FOA) sound encoding. The basic idea of the method is to apply some transformations to the FOA channels (and corresponding labels) to modify and simulate a new DOA of the recorded sounds in a predictable way. Such transformations are: channel swapping and inversion, application of a rotation formula (i.e. Rodrigues' rotation formula) and multiplication by an orthonormal matrix, which correspond to rotations and reflections of the sound sources positions with respect to a reference system centered on the listener.

\section{First Order Ambisonics}
\label{sec:foa}

First-Order Ambisonic (FOA) is a digital audio encoding which describes a soundfield \cite{ambisonics}. It has origin in the B-Format, which encodes the directional information on four channels $W, X, Y$ and $Z$ \cite{ambisonics}. $W$ carries omnidirectional information, while channels $X, Y$ and $Z$ carry the directional information of the sound field along the Cartesian axes of a reference system centered on the listener\cite{ambisonics}.

Adopting the same notation and convention of the dataset used for the following experiments \cite{dataset}, the spatial responses (steering vectors) of the FOA channels are
$H_1\left( \phi, \theta, f \right) = 1$,
$H_2\left( \phi, \theta, f \right) = \sqrt{3}*\sin{\phi}*\cos{\theta}$,
$H_3\left( \phi, \theta, f \right) = \sqrt{3}*\sin{\theta}$, and 
$H_4\left( \phi, \theta, f \right) = \sqrt{3}*\cos{\phi}*\cos{\theta}$,
where $\phi$ and $\theta$ are the azimuth and elevation angles of a sound source, $f$ is frequency and $*$ is used for the multiplication operation. As it is noticeable from the expressions, FOA channels can be seen as the projections of the sound sources to the three dimensional Cartesian axes, with $H_1$ corresponding to channel $W$, $H_2$ to channel $Y$, $H_3$ to channel $Z$ and $H_4$ to channel $X$. Thus, indicating with $\mathbf{S} = \left\{ S_1, ..., S_n \right\}$ a set of sound sources in their STFT domain, FOA channels 
can be written as a sum of each source and its steering vector, that is 
$X = \frac{1}{N} \sum_{n=1}^N H_4(\phi_n, \theta_n, f) * S_n$,
where $N = |\mathbf{S}|$, and
$\phi_n$ and $\theta_n$ are the azimuth and elevation of $S_n$, respectively.

\section{FOA Domain Spatial Augmentation}
\label{sec:prop}
The goal of the method is, from the audio recordings in the dataset, to generate new ones with different DOA information. More specifically, the problem consists in simulating a new set of spatial responses $\{H_i (\phi_n', \theta_n', f) \}_{i=2}^4$ corresponding to new DOA labels $\{\phi_n', \theta_n'\}_{n=1}^N$ for the audio recordings by applying a transformation directly to the FOA channels. It is a known property of FOA that, since it encodes a soundfield rather than the sources themselves, it is possible to apply some operations directly to the channels \cite{ambisonics}, such as rotations and reflections. There are several ways to apply these transformations, leading to different augmentation strategies with different pros and cons. In the following, three strategies are proposed and compared. 

\subsection{First method: 16 patterns}
\begin{table*}[ttt]
\centering
\caption{Sixteen patterns of simple spatial augmentation. 
Swap$(X,Y)$ denotes $X' \gets Y$ and $Y' \gets X$.}
  \begin{tabular}[width=2*\columnwidth]{r|c|c|c|c} \Hline
  				& $\phi - \pi/2$		& $\phi$ 	& $\phi + \pi/2$	& $\phi + \pi$\\  \hline
  $\theta$		& Swap$(-X,Y)$			& original		& Swap$(X,-Y)$			& Swap$(-X,-Y)$\\
  $-\theta$	    & Swap$(-X,Y)$, $Z' \gets -Z$	& $Z' \gets -Z$		& Swap$(X, -Y)$, $Z' \gets -Z$	& Swap$(-X, -Y)$, $Z' \gets -Z$\\ \hline
  				& $-\phi - \pi/2$	& $-\phi$ 		& $-\phi + \pi/2$		& $-\phi + \pi$\\  \hline
  $\theta$		& Swap$(X, -Y)$		& $Y' \gets -Y$			& Swap$(X,Y)$				& $X' \gets -X$\\
  $-\theta$	& Swap$(-X, -Y)$, $Z' \gets -Z$	& $Y' \gets -Y,Z' \gets -Z$		& Swap$(X,Y)$, $Z' \gets -Z$	& $X' \gets -X, Z' \gets -Z$\\ \Hline
  \end{tabular}
  \label{tab:16patterns}
\end{table*}
The \textit{16 patterns} method simply consists in applying to the data one of the 16 prefixed channel transformations summarized in Table \ref{tab:16patterns}, where $\gets$ indicates an assignment. The basic operations used in this method are channel swapping (e.g. $X' \gets Y, Y' \gets X$) and channel sign inversion (e.g. $Z' \gets -Z$) or a combination between the two. Using this set of operations, it is possible to obtain $8$ rotations about the $z$ axis and $2$ reflections with respect to the $xy: z = 0$ plane, for a total of 16 augmentation patterns (i.e. 15 new patterns plus the original one). The corresponding transformations for the labels are also reported in Table \ref{tab:16patterns}. In particular, the listed transformations correspond to the translations of $+0$, $+\pi$, $+\frac{\pi}{2}$ and $-\frac{\pi}{2}$ of the azimuth angles $\phi$ and $-\phi$ and to the pair of opposites $\phi$ and $-\phi$.


The main advantage of this algorithm, other than it's simplicity and straightforward implementation, is the possibility of it being applied to many pre-computed features, such as logmel magnitude spectrogram or phase spectrogram, since the corresponding transformations in the feature-domain are straightforward to compute (channel swapping maps to the same channel swapping, channel sign inversion maps to identity for magnitude and to a 180 degrees difference for phase). Another advantage is that it is easy to control that mapped angles remain in the same domain as the original ones. For example, in the dataset in use for DCASE2019 Challenge task3, all angles are multiples of 10 degrees and elevation angles range from $-40$ to $+40$ degrees. It is easy to see that the augmented angles maintain the same domain. One more important advantage of this method is that it can be applied independently on the number of the maximum number of overlapping sound sources, which is a complication for the next proposed method.

\subsection{Second method: Labels First}
In \textit{Labels First} method, the basic idea is to first decide the target augmented labels, than to apply a transformation to the data accordingly. The critical aspect of this method is that while for azimuth this is always possible independently on the number of overlapping sources, it isn't the same for elevation. The reason is that when modifying the azimuth coordinates by a fixed amount by means of a rotation, \textit{z-axis} is the common rotational axis for all the sources, while for modifying only the elevation coordinate by a fixed amount by means of a rotation, for each source, an appropriate rotation axis must be selected.

Keeping into consideration this critical aspect, assuming at first to have sound files with non-overlapping sound events, the proposed algorithm for this method is follows. For convenience, it is divided in two steps in which azimuth and elevation are augmented separately. In the first step, at first a random angle $\alpha$ is selected and used to translate, at each time step $t$ with arbitrary range, the azimuth labels:
\begin{algorithmic}[lines]
\State $\alpha \gets \mbox{\texttt{random}}(0, 2\pi)$
\State $\phi'_t \gets \phi_t \oplus \alpha$
\end{algorithmic}
where $\oplus$ here indicates an addition with a wrap-around on the domain $(-\pi, \pi)$, i.e. $(\phi_t + \alpha + \pi) \mbox{ mod } 2\pi - \pi$.
At this point, the rotation matrix around \textit{z-axis} $R_z$ is computed:
\begin{equation}\label{Rot_z}
    R_z =
  \begin{bmatrix}
    \cos{\alpha} & -\sin{\alpha} & 0\\
    \sin{\alpha} & \cos{\alpha} & 0\\
    0 & 0 & 1
  \end{bmatrix},
\end{equation}
and applied to the channels at each time step $t$:
\begin{equation}\label{Aug_FOA}
\mathbf{v}_t' = R_z \mathbf{v}_t,
\end{equation}
where
$\mathbf{v}_t = \left(X_t, Y_t, Z_t\right)^{\top}$ denotes original channels and 
$\mathbf{v'}_t = \left(X'_t, Y'_t, Z'_t\right)^{\top}$ denotes azimuth-augmented channels.
In the second step, the elevation coordinate is augmented. At first, a random augmentation angle $\beta$ is selected. To do so, elevation labels in the selected time range (e.g. a batch) is inspected and maximum and minimum values $M_e$ and $m_e$ are extracted. The elevation angle, by definition, has range $\left(-\frac{\pi}{2}, \frac{\pi}{2}\right)$, but in some datasets like the one in use for DCASE2019 Challenge task3, it might have a custom range $(m_{er}, M_{er})$. In order not to go out of this range, the augmentation angle $\beta$ is extracted randomly in the range $(m_{er} - m_e, M_{er} - M_e)$\footnote{In order to maximize the augmentation range, one could segment the audio recordings in frames containing the single sources. Alternatively, one could accept to extend the elevation domain of the dataset and agnostically select a fixed range for the augmentation angle $\beta$, which is convenient when augmenting an entire audio file altogether, as done in experiment 2.}. At this point, elevation labels are updated:
\begin{algorithmic}
\State $\beta \gets \mbox{\texttt{random}}(M_{er} - M_e, m_{er} - m_e)$
\State $\theta'_t \gets \theta_t + \beta$
\end{algorithmic}
Secondly, at each time step $t$, the rotation axis for augmenting elevation is computed. This axis is defined by the unit vector perpendicular to the one along the azimuthal axis, oriented properly so that a rotation of the audio channels by an angle $\beta$ corresponds to the same increment of the elevation label. It can be easily verified with the right-hand rule that this unit vector corresponds to the azimuthal one rotated by $-\frac{\pi}{2}$ about the \textit{z-axis}:
\begin{equation}
    \mathbf{u}_t =
    \begin{pmatrix}
      0 & 1 & 0 \\
      -1 & 0 & 0 \\
      0 & 0 & 1
    \end{pmatrix}
    \begin{pmatrix}
      \cos{\phi'_t} \\
      \sin{\phi'_t} \\
      0
    \end{pmatrix}
    =
    \begin{pmatrix}
      \sin{\phi'_t} \\
      -\cos{\phi'_t} \\
      0
    \end{pmatrix},
\end{equation}
where the first term means $R_z\left(-\pi / 2 \right)$.
Now, Rodrigues' rotation formula is applied to the $\mathbf{v'}_t$, 
we obtain full-augmented channels:
\begin{equation}\label{rodrigues}
    \mathbf{v''}_t = \mathbf{v'}_t \cos{\beta} + (\mathbf{u}_t \times \mathbf{v'}_t)\sin{\beta} + \mathbf{u}_t ~(\mathbf{u}_t \cdot \mathbf{v'}_t) (1 - \cos{\beta}),
\end{equation}
where $\times$ and $\cdot$ denote the cross-product and the inner-product, respectively.

The main advantage of this method is the high control over the augmented labels. For example, it allows for generating new labels which belong to the same domain of the original ones (e.g only multiple of $10^{\circ}$ and elevation restricted to the range $\left(-40^{\circ}, 40^{\circ}\right)$, such as in the dataset used for the experiments\cite{dataset}. The main disadvantage is that it is best suitable for non-overlapping sound events. There are some workarounds to adapt it to sound recordings with multiple overlapping sound events, though. Some possibilities are to apply it only to time frames with one event, to check for the somewhat rare cases in which all of time events share the same azimuth coordinates or to apply a hybrid strategy such as applying the 16 patterns method only for elevation augmentation or considering only one of the overlapping sources and computing labels for the others sources as in Channels First.

\subsection{Third method: Channels First}
Channels first is the most general case of FOA Domain Spatial Augmentation. This method doesn't depend on the number of overlapping sources, but the control over labels is almost completely lost.

The procedure is as follows. A random $(3 \times 3)$ orthonarmal matrix $R$ is selected. An orthonormal matrix $R$ is a matrix such that
$H H^{\top} = I$ and $det(H)=\pm 1$.
This can be done by selecting a random $(3 \times 3)$ matrix and then orthonormalizing it with the Graham-Schmidt method.
Augmented channels 
$\mathbf{v'}$ 
are then computed as:
\begin{equation}\label{Rdot}
    \mathbf{v'} = R ~\mathbf{v}.
\end{equation}
The same transformation is also applied to the labels $y = (\phi, \theta)^{\top}$, in Cartesian coordinates\footnote{Since distance from the listener is not relevant for the task, when converting to and from cartesian coordinates, we always assume the norm $r=1$, that is we consider direction of arrivals as points on the unit sphere.}:
\begin{algorithmic}
\State $y_c \gets \mathrm{to\_cartesian}(y)$
\State $y'_c \gets R ~y_c$
\State $y' \gets \mathrm{to\_spherical}(y'_c)$
\end{algorithmic}
An orthonormal matrix expresses a general rotoreflection. This method allows generating the most number of augmentation patterns for any number of sources, but, since there is few to none control over the labels, it is recommended to use only with datasets without any restrictions on the labels' domain, as justified by the results of experiment 2.

\begin{figure*}[ttt]
  \centering
  \includegraphics[width=179mm,clip]{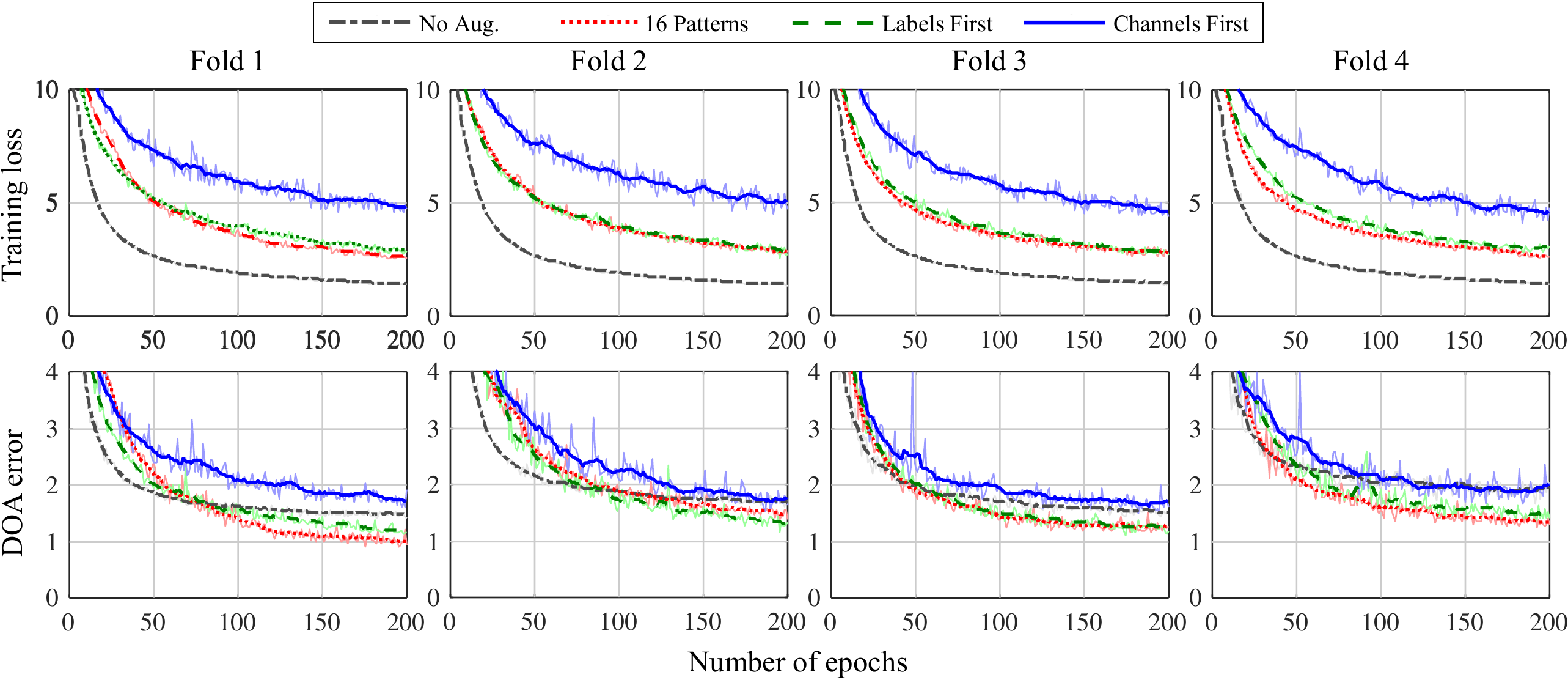}
  \caption{Training progress graph of experiment 2; training loss (top) and DOA error of validation set (bottom). 
  It is apparent that the 16 Patterns method and the Labels First method performed better than without augmentation. The Channels First method lead to worse results, supposedly due to the over-extension of the labels domain and the consequent complication of the problem.}
  \label{fig:training}
\end{figure*}

\section{Experiment}
\label{sec:exp}

\subsection{Experimental setup}
We conducted our experiments, referred to as \textit{Experiment 1} and \textit{Experiment 2}, using two different DOA networks, one simpler, one more sophisticated, here referred to as \textit{Simple DOAnet} and \textit{Sophisticated DOAnet}. 
Both networks give as output a single pair of azimuth and elevation angles computed in a regression fashion and a sound activity detection value computed in a classification fashion. 
Both networks are trained using the maximum overlapping 1 audio files of the DCASE2019 Challenge dataset \cite{dataset}, and evaluated on DOA error (Er) and Frame-recall (FR). We used only overlap 1 files in order to be able to evaluate the effectiveness of FOA Domain Spatial Augmentation specifically for the DOA estimation task. In systems that are able to localize more than one overlapping source, such as SELDnet \cite{SELDnet}, other tasks, such as Sound Event Detection (SED), might be influenced by the augmentation strategy and at the same time influence the performance on DOA estimation.

\subsubsection{Experiment 1}
\textit{Simple DOAnet} has a convolutional recurrent neural network (CRNN) as a core structure, as in \cite{MazzonYasuda, DOAnet, SELDnet, CaoIqbal, Kapka}. Input features are logscale Mel-magnitude spectrogram (logmels) and Generalized Cross-Correlation Phase Transform (GCC-PHAT) of the mutual channels, as in \cite{MazzonYasuda, CaoIqbal}. 
All wav-files were downsampled at a sampling rate of 32 kHz. 
The length of the short-time-Fourier-transform (STFT) and its shift length were 1024 and 640 (20 ms) points, respectively.
The dimension of Mel bins for logmels and GCC-PHAT was 96.
The DNN structure is a CRNN, similar to a SELDnet \cite{SELDnet} without the SED branch and with a single class DOA output.
The CRNN consists of 3 convolutional neural network (CNN) layers, 2 gated recurrent unit layers, and 2 fully-connected (FCN) layers, with the total number of parameters of 545K.

As a loss function, we compute the mean average errors (MAE) between true and predicted labels for both azimuth and elevation and mask them with the true sound activity labels, then sum them to the binary cross-entropy loss of the sound activity output.
The model is trained adopting the four cross-validation folds defined in \cite{dataset} for $400$ epochs each and selecting the best model among the epochs according to the best validation loss. 
The conducted experiments on this model are 3: the first is without using FOA Domain Spatial Augmentation (No Aug), the second is applying the Labels First method on $50\%$ of the input data (LF Half) and the third is applying the Labels First method on all of the input data (LF Full). Augmentation is applied on minibatches of 100 STFT frames ($2\mathrm{s}$).

\subsubsection{Experiment 2}
\textit{Experiment 2} is conducted on \textit{Sophisticated DOAnet}. 
\textit{Sophisticated DOAnet} is a combination method of parametric-based and DNN-based DOA estimation \cite{Yasuda_Dcase_2019}.
Sound intensity vector (IV)-based DOA estimation is used as a base method and two CRNNs are used for denoising and dereverberation of IVs.
Each CRNN consists of 5 CNN layers, 2 FCN layers, and 1 bidirectional long short-term memory layer, and the total number of parameters of \textit{Sophisticated DOAnet} is 2.79M. 
The details of \textit{Sophisticated DOAnet} are described in \cite{Yasuda_Dcase_2019}.

Training was performed on the standard cross-validation folds, and selecting the best model among the epochs according to the best validation loss. 
Four different runs of the training are performed on this network, one without augmentation (No Aug) and one for each of the methods described in Section \ref{sec:prop}: 16 Patterns (16P), Labels First (LF) and Channels First (ChF). 
Based on the results of \textit{Experiment 1}, all data augmentation was performed on $50\%$ of the input data directly on the full length wav files. 
For the Labels First method, elevation augmentation angle $\beta$ was selected randomly between $-20^{\circ}$ and $20^{\circ}$, extending the range of elevation to $(-60^{\circ}, 60^{\circ})$. 

\subsection{Results}

\begin{table}[t]
  \centering
  \caption{Results of \textit{experiment 1} on \textit{Simple DOAnet}}
\begin{tabular}[width=\columnwidth]{ll|c|c|c|c|c}
\Hline
        &       & Fold 1    & Fold 2    & Fold 3    & Fold 4    & Ave.\\ \hline
No      & Er    & 5.32      & 4.85      & 5.56      & 5.07      & 5.22 \\
Aug.    & FR(\%)& 97.78     & 98.46     & 97.79     & 97.67     & 97.93 \\ \hline
LF      & Er    & {\bf 3.34}& {\bf 3.28}& {\bf 3.27}& {\bf 3.07}& {\bf 3.22} \\
Half    & FR(\%)& 98.16     &{\bf 98.89}& 98.28     & 98.14     & 98.37 \\ \hline
LF      & Er    & 3.53      & 3.64      & 3.53      & 3.21      & 3.48 \\
Full    & FR(\%)&{\bf 98.18}& 98.74     &{\bf 98.41}&{\bf 98.38}&{\bf 98.43}\\ 
\Hline
\end{tabular}
\label{tab:MazzonDOAnetResults}
\end{table}

\begin{table}[t]
  \centering
  \caption{Results of \textit{experiment 2} on \textit{Sophisticated DOAnet}}
\begin{tabular}[width=\columnwidth]{ll|c|c|c|c|c}
\Hline
                        &       & Fold 1    & Fold 2    & Fold 3    & Fold 4    & Ave.\\ \hline
No                      & Er    & 1.69      & 1.53      & 1.81      & 1.60      & 1.66 \\
Aug.                    & FR(\%)& 96.91     & 96.46     & 97.14     & 97.50     & 97.00 \\ \hline
\multirow{2}{*}{16P}    & Er    & {\bf 0.96}& {\bf 1.30}& {\bf 1.45}& {\bf 1.21}& {\bf 1.21} \\
                        & FR(\%)& 97.30     & 97.00     & 97.53     &{\bf 97.70}& {\bf 97.39} \\ \hline
\multirow{2}{*}{LF}     & Er    & 1.31      & 1.39      & 1.49      & 1.43      & 1.40 \\
                        & FR(\%)&{\bf 97.34}& 94.16     &{\bf 97.61}&97.14      & 96.56 \\ \hline
\multirow{2}{*}{ChF}    & Er    & 1.99      & 1.35      & 1.98      & 1.61      & 1.73 \\
                        & FR(\%)& 96.45     &{\bf 97.28}& 97.24     & 96.96     & 96.98 \\
\Hline
\end{tabular}
\label{tab:KoizumiDOAnetResults}
\end{table}

\textit{Experiment 1} has mainly two purposes: demonstrate the effectiveness of FOA Domain Spatial Augmentation on training a simple DOA network and discover whether it is best to apply augmentation to all the data or, heuristically, to only half of the data. As reported in Table \ref{tab:MazzonDOAnetResults}, both runs with the use of augmentation outperformed the run without using augmentation on all the cross-validation folds, decreasing the DOA error by $2^{\circ}$ on average and increasing the Frame Recall of $0.5\%$ on average. DOA error achieved the best results by augmenting $50\%$ of the input data ($0.24^{\circ}$ better on average with respect to $100\%$), while augmenting all of the input data achieved the best result in terms of Frame Recall ($0.06\%$ better on average), although the results of these two runs were very close to each other.

\textit{Experiment 2} has the purpose of comparing the different FOA Domain Spatial Augmentation methods illustrated in Table \ref{sec:prop} with each other as well as with non augmented data. The results reported in Table \ref{tab:KoizumiDOAnetResults} show that in this case the 16 Patterns one was the best performing method, followed by the Labels First method, also scoring better than without augmentation in terms of DOA Error. As we expected Labels First to be the method achieving the best scores, we believe the penalty with respect to the expectation is due to the expansion of the labels domain, which means a more difficult problem to solve. As expected, the Channels First method was the least effective on this dataset, scoring worse with respect to non augmented data. It is safe to say that the determining factor for the underperformance is the too big of a difference in the labels domains of the augmented data and of the original data.
In terms of frame recall, again 16 Patterns achieve the best score, although there aren't any particularly noticeable differences.
In Figure \ref{fig:training}, the training progress graphs of experiment 2 are reported. It can be clearly seen that in all the cross-validation folds there is a point since which DOA error on validation split is better with the 16 Patterns and Labels First methods rather than without augmentation.



\section{Conclusions}
\label{sec:concl}
In this paper, FOA Domain Spatial Augmentation, a novel data augmentation strategy, has been proposed. 
The basic idea of the method is to apply rotational transformations to the FOA channels and corresponding labels. 
We proposed three types of such transform: channel swapping and inversion, application of a rotation formula, and multiplication by an orthonormal matrix.
It has been proven effective for training two different neural networks for the task of DOA estimation, improving the DOA error by $40\%$. Future research will be to further investigate the effectiveness of this augmentation strategy in different scenarios, for example with a dataset including all the possible DOAs or with overlapping sound events.

%

\end{sloppy}
\end{document}